\documentclass{uai2022} 

\usepackage[american]{babel}

\usepackage{natbib} 
\usepackage{mathtools, amsmath, amssymb, colonequals} 
\usepackage{amsthm}
\usepackage{xcolor}
\usepackage{paralist}

\theoremstyle{definition}
\newtheorem{example}{Example}
\newtheorem{definition}{Definition}

\usepackage{booktabs} 
\usepackage{tikz} 
\usetikzlibrary{calc,positioning,arrows,shapes,fit,backgrounds,scopes}
\usepackage{multirow}

\title{Inductive Synthesis of Finite-State Controllers for POMDPs}

\usepackage{multirow}

\newcommand{\sinit}{s_{0}}
\newcommand{\Act}{Act}
\newcommand{\Obs}{Z}
\newcommand{\obs}{z}
\newcommand{\obsFun}{O}
\newcommand{\node}{n}
\newcommand{\nodes}{N}
\newcommand{\mpm}{P}
\newcommand{\mdpT}{(S,\sinit,\Act,\mpm)}
\newcommand{\mdp}{M}
\newcommand{\fscT}{(\nodes,\node_0,\gamma,\delta)}
\newcommand{\fsc}{F}
\newcommand{\ffsc}{\mathcal{F}}
\newcommand{\abst}[1]{\mathcal{A}^{#1}}

\newcommand{\pomdpT}{(S,\sinit,\Act,\mpm,\Obs,\obsFun)}
\newcommand{\pomdp}{\mathcal{M}}
\newcommand{\imc}{\pomdp^\fsc}
\newcommand{\ifmc}{\pomdp^\ffsc}

\newcommand{\iverson}[1]{\ensuremath{[#1]}}

\newcommand{\policy}{\pi}

\newcommand{\given}{~\vert~}
{}

\newcommand{\unitinterval}{[0,1]}

\newcommand{\prob}{P}
\newcommand{\rew}{R}

\author[1]{Roman Andriushchenko}
\author[1]{Milan \v{C}e\v{s}ka [\textrm{ceskam@fit.vutbr.cz}]}
\author[2]{Sebastian Junges}
\author[3]{Joost-Pieter Katoen}
\affil[1]{%
    Brno University of Technology, Brno, Czech Republic\\
}
\affil[2]{%
    Radboud University, Nijmegen, Netherlands\\
}
\affil[3]{%
    RWTH Aachen University, Aachen, Germany
  }
  
\begin{document}

\maketitle

\begin{abstract}
\vspace{-1em}
 We present a novel learning framework to obtain finite-state controllers (FSCs) for partially observable Markov decision processes and illustrate its applicability for indefinite-horizon specifications.
Our framework builds on oracle-guided inductive synthesis to explore a design space compactly representing available FSCs. The inductive synthesis approach consists of two stages: The outer stage determines the design space, i.e., the set of FSC candidates, while the inner stage efficiently explores the design space. 
This framework is easily generalisable and shows promising results when compared to existing approaches.
Experiments indicate that our technique is (i) competitive to state-of-the-art belief-based approaches for indefinite-horizon properties, (ii) yields smaller FSCs than existing methods for several models, and (iii) naturally treats multi-objective specifications.
\end{abstract}

\vspace{-1em}
\section{Introduction}\label{sec:intro}
 Partially observable MDPs (POMDPs) model sequential decision making processes in which the agent only observes limited information about the current state of the system~\citep{smallwood1973optimal,kaelbling1998planning}. The key challenge in the analysis of POMDPs is to compute a policy satisfying some constraints, captured as a threshold on (discounted) reward or as a task description given in, e.g., a temporal logic. In full generality, policies need arbitrary memory to reflect the belief state of the agent. Point-based~\citep{DBLP:journals/jair/PineauGT06,DBLP:journals/jair/SpaanV05} and Monte Carlo methods~\citep{silver2010monte} excel in finding such policies. Solving the generally undecidable policy learning problem profits from having complementary approaches in the portfolio. A natural alternative is to search for (small) finite state controllers (FSCs)~\citep{hansen1998solving}. Such controllers provide benefits in terms of explainability~\citep{bonet2010automatic,wang2019state},  resource-consumption~\citep{grzes2013isomorph}, and generalisability~\citep{DBLP:conf/iclr/InalaBTS20}. Recently, an automata learning framework has been proposed for synthesising permissive FSCs~\citep{Bo2021Supervised}. In this paper, we propose a novel approach---\emph{inductive synthesis}---to find FSCs for POMDPs.
 
Inductive synthesis is a technique developed in the context of program synthesis, originally proposed by Church in the 1950's, the task to construct a program that provably satisfies a given formal specification. As developing a program (or in this context, a controller) from scratch is mostly infeasible, variants emerged, most notably syntax-guided synthesis~\citep{DBLP:series/natosec/AlurBDF0JKMMRSSSSTU15-short,DBLP:journals/cacm/AlurSFS18} variations such as \emph{sketching}~\citep{DBLP:conf/asplos/Solar-LezamaTBSS06}.
In sketching, the user provides a sketch that outlines a controller implementation, and a specification that constrains the controller's behaviour. 
The principal engine behind many instances of sketching is (oracle-guided) \emph{inductive synthesis}~\citep{DBLP:journals/acta/JhaS17} and falls in a more general framework of learner-teacher frameworks. 
In a nutshell, this methodology suggests to heuristically \emph{guess} candidate solutions, to \emph{validate} them, and in case the solution is not satisfactory, \emph{learn} in order to refine the search heuristic. 
The successful application of inductive synthesis has inspired numerous applications beyond classical programming, including recent works on sketching of probabilistic programs~\citep{DBLP:conf/pldi/NoriORV15,DBLP:journals/fac/0002HJK21,DBLP:conf/cav/AndriushchenkoC21} and (variations of) programmatic reinforcement learning~\citep{DBLP:conf/icml/VermaMSKC18,DBLP:conf/iclr/InalaBTS20}. 
\emph{This paper proposes inductive synthesis to search for FSCs in POMDPs.}



\begin{figure}
    \centering
    \begin{tikzpicture}
        \node[inner sep=4pt,draw] (lout) {Learner};
        \node[inner sep=4pt,right=2cm of lout,draw,fill=black!5] (tout) {Teacher};
        
        \node[below=1.2cm of lout,inner sep=4pt, draw] (lin) {Searcher};
        \node[below=1.2cm of tout,inner sep=4pt, draw] (tin) {Eval};
        \node[left=2.6cm of lout,yshift=-2.8em,draw, minimum height=3cm] (abstr) {\rotatebox{90}{Abstr Oracle}};
        
          \draw[->] (lin) edge[bend left=10] node[above] (f) {\footnotesize{FSC}} (tin);
        \draw[->] (tin) edge[bend left=10] node[below] (p) {\footnotesize{value $\&$ conflicts}} (lin);
        
        \node[fit=(lin)(tin)(p)(f), inner sep=2pt, dashed, draw] (teacherdet) {};
        {[on background layer]
        \draw[dotted] (tout.north west) -- (teacherdet.north west);
        \draw[dotted] (tout.south east) -- (teacherdet.south east);
        
        \draw[dotted] (tout.south west) -- (teacherdet.south west);
        \draw[dotted] (tout.north east) -- (teacherdet.north east);
          
        \node[fit=(lin)(tin)(p)(f), inner sep=2pt, dashed, draw, fill=black!5] (teacherdet) {};
        }

        \draw[->] (lout) edge[bend left=10] node[above,fill=white] {\footnotesize{design space}} (tout);
        \draw[->] (tout) edge[bend left=10] node[below,fill=white] {\footnotesize{best FSC}} (lout);
        
        \draw[->] (lout) edge[bend right=10] node[above] {\footnotesize{design space}} (abstr.76);
        
        \draw[<-] (lout) edge[bend left=10] node[below] {\footnotesize{value bounds}} (abstr.71);
        
        \draw[->] (lin) edge[bend right=10] node[above] {\footnotesize{(sub)design space}} (abstr.-64);
        
        \draw[<-] (lin) edge[bend left=10] node[below] {\footnotesize{value bounds}}  (abstr.-72);

    \end{tikzpicture}
    \caption{Nested Inductive Synthesis Framework with an Abstraction Oracle. The framework takes a POMDP and a specification and finds an FSC that satisfies the specification.}
    \vspace{-1.1em}
    \label{fig:overview}
\end{figure}
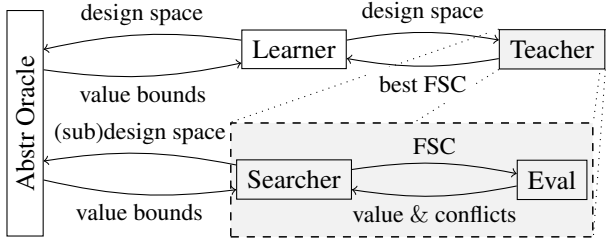
Our inductive synthesis framework works in two stages, see Fig.~\ref{fig:overview}. 
Let us first discuss the outer stage.
Here, a learner constructs a \emph{design space} containing (finitely many) FSCs. A teacher provides the `best' FSC within this design space, and potentially additional diagnostic information. The learner either accepts the FSC provided by the teacher as final result, or adapts the design space. Naturally, teachers will provide much better FSCs much faster whenever the design space for these FSCs is small. The key ingredient for the outer stage is thus to start with a small design space and to strategically adapt this design space based on the obtained feedback from the teacher. A similar idea was proposed in~\citep{kumar2015history}, where the entropy of the observations is used as criterion for adding memory to the FSC. We use the FSC returned by the teacher together with the state-values induced by this FSC. Additionally, we use an abstraction oracle, see below.


The inner stage describes the internals of the teacher that determines the `best' FSC within the design space. 
The teacher may naively use enumeration, but can also be realised using  branch-and-bound~\citep{grzes2013isomorph} or mixed-integer linear programming (MILP)~\citep{DBLP:conf/aaai/AmatoBZ10,kumar2015history}. 
We realise the teacher by (another) inductive synthesis loop.
We search for an FSC by symbolically representing the design space as a propositional logic formula.
The policy evaluation analyses the fixed policy w.r.t.\ a given specification (e.g., a reward function and a threshold). 
If the policy refutes the specification, the evaluation engine indicates the distance to satisfaction (e.g., the achieved value) as well as conflicts---critical parts of the FSC that suffice to violate the specification---that are used to prune the search design  space~\citep{DBLP:journals/fac/0002HJK21}. 

Both learning stages have access to an additional oracle that, inspired by ~\cite{andriushchenko2021inductive}, \emph{over-approximates} the design space. 
This larger abstract design space can efficiently be analysed as the underlying problem solved by the oracle resembles the analysis of fully observable policies. The oracle yields constraints to what the best FSC within the original design space will possibly achieve.
This information is an essential ingredient to guide the search in both stages.


The separate policy evaluation---a natural component in an inductive synthesis framework---brings some advantages.
The policy evaluation (i.e., solving systems of linear equations) via dedicated algorithms is faster than letting an (MI)LP solver solve these equations~\citep{DBLP:conf/atva/DehnertJWAK14}. This improves upon performance of MILP-based approaches (either primal \citep{winterer2020strengthening} or dual \citep{kumar2015history}) for  FSC synthesis.
Furthermore, as the policy is fixed, our framework provides an elegant alternative to existing approaches for constrained POMDPs~\citep{poupart2015approximate,khonji2019approximability} and multi-objective POMDPs~\citep{soh2011evolving,roijers2013survey,wray2015multi}.
It additionally paves the way to learn robust FSCs for POMDPs with imprecise probabilities, similar to~\citep{DBLP:conf/aaai/Cubuktepe0JMST21}. 

We instantiate our framework to learn \emph{deterministic} FSCs, i.e., FSCs that do not use randomisation.
Finding optimal deterministic FSCs is NP-complete whereas finding optimal randomised FSCs is ETR-complete\footnote{The class ETR lies between NP and PSPACE.}~\citep{junges2018finite,DBLP:journals/jcss/JungesK0W21}. 
Algorithmically, finding randomised FSCs requires solving non-convex optimisation problems with thousands of variables.
This often limits the guarantees on global (almost-)optimality that are practically feasible~\citep{kumar2015history}. 
Deterministic FSCs are additionally beneficial in terms of reproducibility of their behaviour, which is useful for debugging.
We use an evaluation framework that supports indefinite horizon queries, e.g., queries with a discount factor one. These support queries with a discount factor, but also settings as used in Goal-POMDPs~\citep{DBLP:conf/ijcai/BonetG09,DBLP:conf/aips/KolobovMWG11}.
These queries naturally occur when using  temporal logic specifications and are particularly adequate for safety-critical aspects.





The experimental evaluation shows the applicability of our approach on a wide range of benchmarks with promising results. Particularly, it significantly outperforms approaches based on MILP optimisation. 
We further compare it with the state-of-the-art belief-based approaches, namely, with recent works in formal verification on under-approximation for indefinite-horizon specifications~\citep{norman2017verification,bork2022under}. 
Our inductive synthesis approach is highly competitive and for several POMDPs (having a moderate number of observations/actions and large/infinite belief-space), it is able to find small FSCs improving lower bounds of existing solutions.

\section{Problem Statement}
A \emph{(discrete) distribution} over a finite set $X$ is a~function $\mu \colon S \rightarrow \unitinterval$~s.t.~$\sum_x \mu(x) = 1$. The set $Distr(X)$ contains all distributions over $X$. 

  A \emph{Markov decision process (MDP)} is a tuple $\mdp=\mdpT$ with a finite set $S$ of \emph{states}, an initial state $\sinit \in S$, a finite set $\Act$ of \emph{actions}, and a \emph{transition probability function} $\mpm(s' \given s,a)$ that gives the probability of evolving to $s'$ after taking action $a$ in $s$.
 A \emph{Markov chain} (MC) is an MDP with $|\Act| = 1$; its transition function is written as $\mpm(s' \given s)$.
Markov decision processes can additionally be equipped with a reward function $r(s,a)$. We do not use discount factors, see the paragraph on specifications below.



A \emph{Partially Observable MDP (POMDP)} $\pomdp = \pomdpT$ extends MDP $\mdp$ with  a finite set $\Obs$ of \emph{observations}, and  a (deterministic) \emph{observation function} \footnote{Observation functions resulting in a distribution over observations can be encoded by deterministic observation functions at the expense of a polynomial blow-up~\citep{ChatterjeeCGK16}.}
$\obsFun$ that returns for every state $s$ an observation $\Obs(s) = \obs \in \Obs$. The observation $z \in \Obs$ is said to be \emph{trivial} if there is only one state $s\in S$ with $\obsFun(s) = z.$ 

\textbf{Finite State Controllers (FSCs)}
are automata that compactly represent policies. We call its states nodes to distinguish them from POMDP states. 
FSCs in the literature come in various styles, in particular either as Moore machines, with the output---the action it selects---determined by the node, or as Mealy machines, with the output determined by the taken transition~\citep{DBLP:conf/aaai/AmatoBZ10}.
In the context of sketching FSCs and their inductive exploration, it is convenient to describe controllers as Mealy machines. 
Furthermore, we restrict ourselves to deterministic FSCs. 


Formally, a \emph{finite-state controller} (FSC) for a POMDP $\pomdp$ is a tuple $\fsc = \fscT$, where $N$ is a finite set of \emph{nodes}, $\node_0 \in N$ is the \emph{initial node}, $\gamma(\node,\obs)$ determines the action when the agent is in node $\node$ and observes $\obs$, while $\delta$ updates the memory node to $\delta(\node,\obs)$, when being in $\node$ and observing $\obs$. 
For $|\nodes|=k$, we call an FSC a $k$-FSC. 

\begin{figure}
    \centering
    \includegraphics[width=0.24\textwidth]{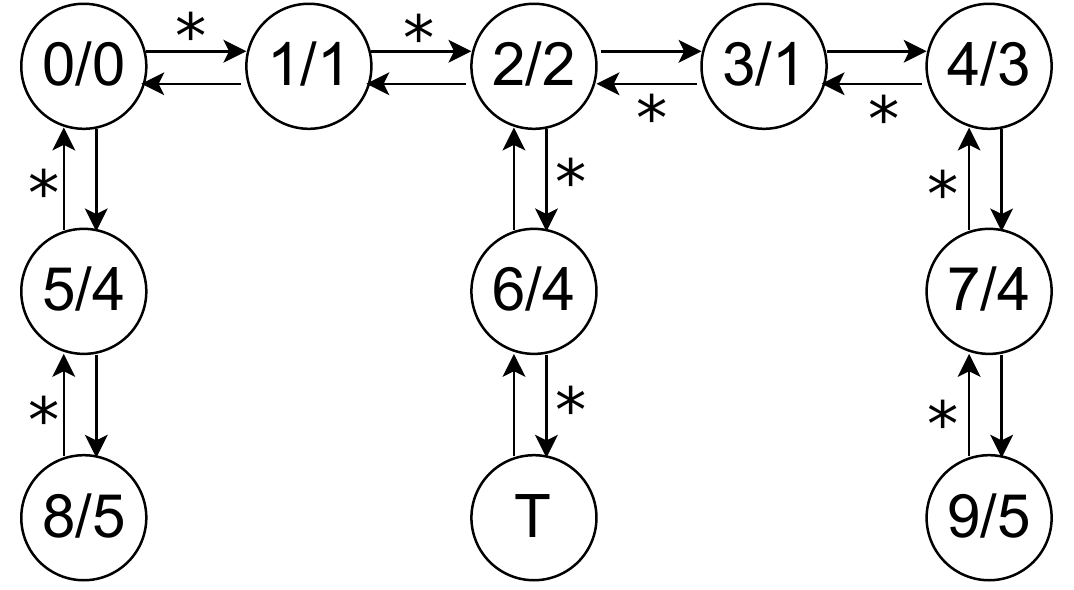} 
    \includegraphics[width=0.2\textwidth]{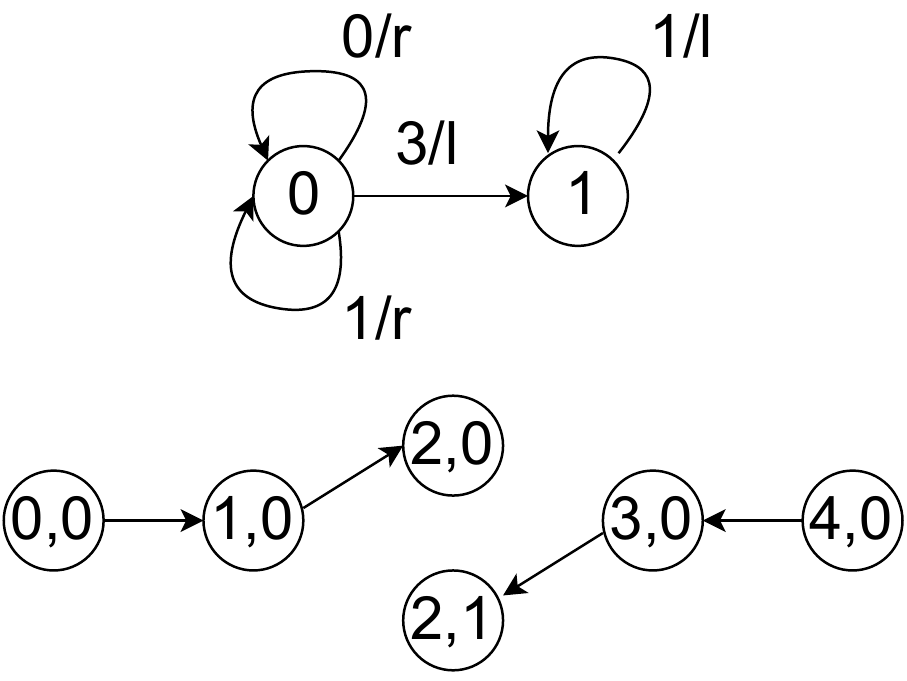}
    \caption{A simple maze problem (left), a part of a 2-FSC (right, top) and a part of the induced MC (right, bottom).}
    \label{fig:maze_1}

\end{figure}

Imposing $k$-FSC $\fsc$ onto POMDP $\mathcal{M}$ yields the Markov chain (MC) $\imc = (S^{\fsc} , (s_0,n_0), \mpm^{\fsc})$ with ${S^{\fsc} = S \times N}$ and using\footnote{ Iverson-brackets: $\iverson{x} = 1$ if predicate $x$ is true, $0$ otherwise.} $z = \obsFun(s)$:
$$\mpm^{\fsc}((s',n') \given (s,n)) = \mpm (s' \given s, \gamma(n,z)) \cdot \iverson{\delta(n,z) = n'}.$$

\begin{example}
As running example, we use a simple variant of the maze problem~\citep{hauskrecht1997incremental}, where an agent tries to reach the state $s_T$, modelled by the POMDP $\pomdp$ with $S=\{s_0, \ldots s_9, s_T\}$, $\Act=\{u,d,l,r\}$, and $\Obs=\{z_0,\ldots,z_5\}$. The initial state is given by a uniform distribution over~$S$. Fig.~\ref{fig:maze_1}(left) depicts $P$ and $\obsFun$ where state $s_x$ is labelled by $x/y$ with $x$ the state index and $y$ its observation, i.e., $O(s_x) = z_y$. The arrow direction from $x/y$ to $x'/y'$ represents the action; e.g., $\rightarrow$ corresponds to action $r$. The maze is slippery. An action is successful with probability $0.9$; with $0.1$, the agent does not move. Actions without effect are omitted from the figure. Fig.~\ref{fig:maze_1} (right, top) illustrates a fragment of a 2-FSC where $\gamma(0,0) = \gamma(0,1) = r$ (for memory node 0 and observations $z_0$ and $z_1$, action $r$ is chosen), $\gamma(0,3) = \gamma(1,1) = l$, $\delta(0,0) = \delta(0,1) = 0$ (memory node 0 is not changed for $z_0$ an $z_1$) and $\delta(0,3) = \delta(1,1) = 1$. This FSC tries to resolve the inconsistency (formalised later) in the observation $z_1$, i.e., in $s_1$ the action $r$ is optimal but in $s_3$ action $l$ is optimal (w.r.t.\ reaching $s_T$). Fig.~\ref{fig:maze_1} (right, bottom) illustrates a fragment of the induced MC containing two copies of $s_2$.
\end{example}

\textbf{Specifications} 
contain two parts: a set of \emph{constraints} given by quantitative properties and a single optimisation \emph{objective}.
Constraints are defined as indefinite-horizon reachability and expected reward properties, but our approach also supports more general probabilistic temporal logic properties~\citep{BK08}\footnote{These properties can describe thxe setting of  goal-POMDPs, finite horizon reachability and rewards, and discounted rewards.}
Let target set $T\subseteq S$, \emph{thresholds} $\lambda_1 \in \unitinterval$ and $\lambda_2 \in \mathbb{R}^+$ 
and $\bowtie \, \in \{ \leq, \geq\}$. 
The POMDP $\pomdp$ under FSC $\fsc$ \emph{satisfies} the constraint $\prob_{\bowtie \lambda_1}$ if the probability $\Pr^{\fsc}$ of reaching $T$ in the induced MC $\imc$ meets $\bowtie \lambda_1$. 
Similarly, the constraint $\rew_{\bowtie \lambda_2}$ is satisfied if the expected reward $R^{\fsc}$ accumulated in MC $\imc$ until reaching $T$ meets $\bowtie \lambda_2$. We call an FSC $\fsc$ \emph{admissible} (for $\pomdp$), if $\pomdp$ under $\fsc$ satisfies the given (set of) constraint(s).
Objectives either minimise or maximise reachability probabilities (as in goal-POMDPs) or (un)discounted expected reward properties, denoted as $\prob_*$ and $\rew_*$ respectively for $* \in \{\min,\max\}$. The probability or reward obtained by FSC $F$ on $\pomdp$ is called the \emph{value} of $F$.
For conciseness, we assume throughout the paper that the specification contains a maximisation objective. Minimisation is analogously supported (but may require flipping bounds and inequalities).

\paragraph{Problem statement.}
We aim to construct an algorithm that: 
i) quickly finds a (small) admissible FSC $F$ and ii) incrementally improves $F$ w.r.t.\ the optimisation objective. 
We can view the algorithm as solving a sequence of decision problems, where the first decision problem is to find some admissible FSC $F_0$, and decision problem $i{+}1$ is to find an admissible FSC $F_{i+1}$ whose value improves upon the value of the previous FSC $F_i$.

\section{Inductive exploration of FSCs}
\vspace{-0.5em}
\label{sec:innerloop}
This section presents the inner loop (see Fig.~\ref{fig:overview}) in which we search among a given set of $k$-FSCs. Before we describe the ingredients, we formalise the representation of the set of $k$-FSCs. We then outline the two oracles that our search can use to prune the search space. A \emph{hybrid strategy}~\citep{andriushchenko2021inductive} combines the two oracles by switching based on perceived performance while communication between the oracles takes place.
\vspace{-0.5em}
\subsection{Families of FSCs}
\label{sec:red_fsc}

A POMDP and a single FSC induce a Markov chain. A POMDP and a set of FSCs thus induces a set of MCs. The set of FSCs has additional structure enabling a concise representation as a family-MC. We first consider \emph{full} FSCs where for each observation the same amount of memory is used. We generalise this to a class of \emph{reduced} FSCs that are more memory efficient.



\begin{definition}
A \emph{family} of full $k$-FSCs is a tuple $\ffsc_k = (N, n_0, K)$, 
where $N$ is the set of $k$ nodes, $n_0 \in N$ is the initial node, $K = N\times \Obs$ is a finite set of parameters each with domain $V_{(n,\obs)} \subseteq \Act\times N$.
\end{definition}

An FSC is obtained by choosing values for each parameter. Their domains determine the set of FSCs described by the family. Families of FSCs thus contain $\mathcal{O}(|N||\Obs|)$ many FSCs. To simplify the notation, we will 
interpret $\ffsc_k$ as a set of $k$-FSCs. 
A POMDP $\pomdp$ and a family $\ffsc_k$ naturally induces the family of MCs $\ifmc = \{\imc \mid \fsc \in \ffsc_k  \}$.


\begin{example}
The family $\ffsc_2$ of all 2-FSCs for our maze problem is given by $N = \{n_0,n_1\} $, $K = \{ (n_i,z_j) \mid i \in \{0,1\} \wedge  j \in \{u,d,l,r\} \}$, $V_{(n,z)} = \{u,d,l,r\} \times \{n_0,n_1\}$ for all  $(n,z) \in K$.
\end{example}


While FSCs have $k$ available memory nodes in conjunction with every observation, memory is often required only in some observation (see e.g., the running example). Therefore, we consider reduced FSCs given by a \emph{memory model} $\mu: \Obs \rightarrow \mathbb{N}$, where $\mu(\obs)$ determines the number of memory nodes used in the observation $\obs$. The reduced FSC requires $\max_{\obs \in\Obs}\{\mu(\obs)\}$ nodes, but the parameter domains are significantly reduced. In the previous example as well as in various benchmarks, memory is required only in a few observations, dropping the number of mappings to $\mathcal{O}(|N|{+}|\Obs|)$. 

\begin{definition}
 A \emph{reduced family} $\ffsc_{\mu}$ given by the memory model $\mu$ is a sub-family of $\ffsc_k$ for $|N| = \max_{\obs \in\Obs}\{\mu(\obs)\}$ where $(n,z)\in K$ implies $n\leq \mu(z)$, and the domains $V_{(n,z)}$ are as in $\ffsc_{k}$\footnote{If a memory update $\delta(n,z)=n'$ is invalid in the resulting observation $z'$  (i.e. $n'>\mu(z')$), the update to $n_0$ is used.}. 
\end{definition}

Such reduction has several key benefits: in the case of resource-aware applications, the memory needed to store and execute the controller is reduced, better interpretability of the controller is achieved, and finally the family of reduced FSCs induces a smaller design space (the number of parameters is given by the size of the mappings).

\vspace{-0.5em}
\subsection{MDP abstraction}
\vspace{-0.5em}
\label{sec:abst}


As families of MCs may become large, it is beneficial to consider an abstraction (represented as a single MDP) of it. 

\begin{definition}
MDP $\abst{\ffsc} = (S{\times}N,(s_0,n_0), \Act^{\ffsc}, \mpm^{\ffsc})$ is an \emph{abstraction} of MC family $\ifmc$ with $\Act^{\ffsc} = \Act \times N$ and $\mpm^{\ffsc}((s',n') \given (s,n),(a,n')) = \mpm(s' \given s,a) $ if $(a,n')\in V_{(n,\obsFun(s))}$, and 0 otherwise.
\end{definition}

For MDPs and our specifications, it suffices to consider deterministic memoryless policies, i.e., policy $\policy$ for MDP $\abst{\ffsc}$ is a function $\policy\colon S{\times}N \rightarrow \Act$.
It is \emph{consistent} if $\obsFun(s) = \obsFun(s')$ implies $\pi((s,n)) = \pi((s',n))$ for all $s,s' \in S, n\in N$. 
The set of consistent polices in $\abst{\ffsc}$ matches the family $\ffsc$.
The parameter $(n,z) \in K$ is \emph{inconsistent} in $\pi$ if $\exists {s,s'\in S}: \obsFun(s) = \obsFun(s') = z \wedge \pi((s,n)) \neq \pi((s',n))$. 
The observation $\obs \in \Obs$ is inconsistent if the parameter $(n,z) \in K$ is inconsistent for some $n \in N$.


\begin{example}
Assume we want to maximise the probability to reach $s_T$. 
The stars in Fig.~\ref{fig:maze_1} represent the optimal policy~$\pi^{*}$ in MDP $\abst{\ffsc}$ where $\ffsc$ is set of all \mbox{1-FSCs} for the maze problem. $\pi^*$ is inconsistent in the observations $z_1$ and $z_4$.  
\end{example} 

The analysis of MDP $\abst{\ffsc}$ provides useful information about the family $\ffsc$. 
Assume our interest is to maximise the constraint $\prob_{\geq \lambda}$ for target set $T$.
(Reasoning for constraints of the form $\prob_{\leq \lambda}$ is dual and expected reward constraints are handled the same).
Let policy $\pi^{*}$ in $\abst{\ffsc}$ achieve the probability $\Pr^{\pi^{*}}$. 
If $\Pr^{\pi^{*}} < \lambda$, it is guaranteed that all $F \in \ffsc$  violate the constraint and $\ffsc$ can be safely pruned. 
Otherwise, we check the consistency of policy $\pi^{*}$.  
If it is consistent, it represents a valid FSC satisfying the constraint.
Similarly, a minimising policy can be used to prove that the entire family $\ffsc$ satisfies the constraint.
If the analysis of $\abst{\ffsc}$ is inconclusive, $\ffsc$ has to be \emph{refined}.
Additionally, analysing $\abst{\ffsc}$ provides state-vectors $ub$ 
and $lb$ such that  $\forall s \in S$, $ub(s)$ and $lb(s)$ represent the maximal and minimal probability to reach $T$ from $s$, respectively. These bounds are used in the inner and outer synthesis loop.\footnote{Furthermore, the state-vectors $ub$ and $lb$  allow \emph{bootstrapping} the analysis of MDP $\abst{\ffsc_i}$ where $\ffsc_i$ is a subfamily of $\ffsc$: 
This exploits the fact that $\ffsc_i$ shares the structure of $\ffsc$ while some actions for some states are removed.}

For specifications with \emph{multiple} constraints, the optimisation objective is handled by iteratively updating a new (initially trivial) constraint representing the running value of the optimum so far. 
Once a policy $\pi$ satisfying all constraints is found, we update the new constraint according to the objective value that $\pi$ achieves. Reasoning about multiple constraints works as follows. 
If the entire family $\ffsc$ violates some constraint, it is pruned. 
Otherwise, we investigate the consistency of the policies found for the constraints with the aim to find a FCS improving the optimum and eventually to prune $\ffsc$. 
If $\ffsc$ cannot be pruned, it is \emph{refined}.     

\paragraph{Refinement strategy}
The refinement strategy is a key component in driving the exploration of the family $\ffsc$. 
It decomposes $\ffsc$ into sub-families by splitting the domain of selected parameters from $K$. 
In contrast to the general refinement strategy used in program synthesis~\citep{ceska2019shepherding}, we leverage the specific topology of the FSC families. 

The key idea is to examine the inconsistencies of the policies $\pi^*$ obtained for particular constraints\footnote{We focus on the constraint derived from improving the objective value as it usually is the most restrictive.}. 
We estimate the \emph{significance} of each inconsistent parameter $p \in K$ in $\pi^*$ by examining the impact of changing $p$ in $\pi^*$ weighted by the expected frequency of the decisions corresponding to~$p$. 
We then select the most significant parameter~$p$. 
Assume (inconsistent) $p$ has domain $V_p = \{v_1, \ldots, v_n \}$ and $\pi^*$ selected the options $v_i$, $v_j$. 
We partition $V_p$ into $V_p^1 = \{v_i\}$, $V_p^2 = \{v_j\}$ and $V_p^3 = V_p \setminus \{v_i, v_j\}$ and create the three corresponding subfamilies.
This allows us to remove the inconsistency of $p$ by considering the selected options $v_i$ and $v_j$ by different sub-families. 

We can restrict the exploration to FSCs that are structurally close to $\pi^*$. 
We use the \emph{incomplete} refinement strategy that i) fixes the options selected by $\pi^*$ in perfectly observable states, ii) fixes the options in the consistent parameters, and iii) removes options in the inconsistent parameters that were not selected by $\pi^*$, i.e., the set $V^3$.   


\subsection{Counter-example-based pruning}
\label{sec:counterexpruning}
While MDP abstraction is an effective exploration strategy for large FSC families, it can be helpful to prune the design space by analysing candidate FSCs. If the individual candidate is admissible and has a good value, this clearly accelerates the teacher. Otherwise, we learn counterexamples.
To realise this teacher, we represent the FSCs that have not been pruned as a propositional formula. 
We use the SMT solver CVC5~\citep{cvc5} (over quantified-free bounded integers) to effectively manipulate the propositional formula and to find FSCs that have not been pruned.

We assume  a constraint  $\prob_{\geq \lambda}$ (for target set $T$), a family $\ffsc$, the state-vector $ub$ obtained from the maximising policy $\pi^*$ in MDP $\abst{\ffsc}$, and FSC $\fsc \in \ffsc$. 
\begin{definition}
A~\emph{counter-example} (CE) for FSC $\fsc$ and $\prob_{\geq \lambda}$  is a subset $C\subseteq S^{\fsc}$ that induces the sub-MC of $\imc$ given as $({C \cup \mathrm{succ}(C) \cup \{s_{\bot}}, s_{\top} \}, (s_0,n_0), P')$ where $P'(s) =$
\begin{align*}
\begin{cases}
P^{\fsc}(s) & \text{if } s \in C, \\
[s_{\top} \mapsto ub(s),s_{\bot} \mapsto 1{-}ub(s)]  & \text{if } s \in \mathrm{succ}(C) \setminus C, \\
[s \mapsto 1] & \text{if } s \in \{s_{\top}, s_{\bot} \}, \\
\end{cases}
\end{align*}
where $\mathrm{succ}(C)$ is the set of direct successors of $C$, and the probability to reach $T\cup \{s_{\top}\}$ from $C$ is $< \lambda$. 
\end{definition}

 Intuitively, states $s$ outside the CE evolve to $s_{\top}$ with probability $ub(s)$, the maximal probability to reach $T$ from $s$ in the family $\ffsc$ (i.e., the worst that can happen for CE in $\ffsc$).
 They evolve to $s_{\bot}$ with probability $1{-}ub(s)$, the minimal probability to avoid $T$ in $\ffsc$. 
 For $(s,n) \in C$, the parameter $(n,\obsFun(s)) \in K$ is called \emph{relevant}. The CE for the constraint $\prob_{\leq \lambda}$ is defined  similarly using $lb$ rather than $ub$.
 
For each $\fsc' \in \ffsc$ that selects the same options as $\fsc$ for all relevant parameters of the CE $C$, it holds that $P^{\fsc'} < \lambda$. 
Therefore, we can safely remove $\fsc'$ from the design space. 
We say that $C$ \emph{generalised} to the set of all such $\fsc'$. 

Smaller CEs lead to more effective pruning. 
As computing minimal CEs is NP-complete ~\citep{funke2020farkas}, we use a greedy approach proposed in~\citep{andriushchenko2021inductive}. 
Handling multiple constraints is straightforward as we can compute the CE for each constraint violated by the FSC $\fsc$. This can potentially improve the pruning.

Similarly as the incomplete refinement strategy in Sec.~\ref{sec:abst}, we consider an incomplete generalisation of the CEs. 
In particular, we redefine the notation of relevant parameters. 
The parameter $(n,\obsFun(s))$ for $(s,n) \in C$ is relevant only if the observation $\obsFun(s)$ is inconsistent in $\abst{\ffsc}$ or the option selected by $\fsc$ is different from the options selected by $\pi^*$. 
This leads to more aggressive pruning and restricts the exploration to the FSCs that are topologically close to $\pi^*$.

\begin{example}
Consider a variant of our maze problem with initial state $s_0$, family $\ffsc$ of all 1-FSCs, where the available set of actions in the observation $o_3$ is restricted s.t.\ $V_{(n_0,z_3)} \in \{u,d,r\} \times \{0\}$. Let FSC $\fsc$ as in Fig~\ref{fig:CE}~(left). The right part illustrates the induced MC and the middle part shows the CE $C$ for the constraint $\prob_{\geq 1}$.
Note $P((s_4,n_0),s_{\bot}) = 1$ as $ub(s_4)=0$.
Thus, the relevant parameters are $(n_0,z_i)$ for $i\in \{0,1,2\}$. The generalisation of $C$ enables pruning a significant part of $\ffsc$. 
Under incomplete generalisation, the parameter $(n_0,z_0)$ is not relevant as it is consistent in $\abst{\ffsc}$ and $\fsc$ picks the same option as $\pi^*$.  
\end{example}

\begin{figure}
    \centering
     \includegraphics[width=0.47\textwidth]{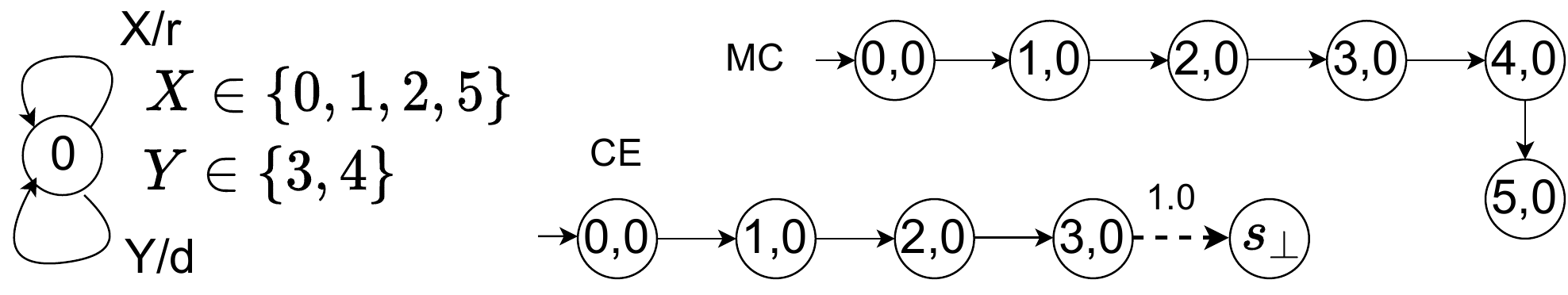} 
    \caption{A CE for the given FSC in the maze problem.}
    \vspace{-1em}
    \label{fig:CE}
\end{figure}

\vspace{-0.5em}
\section{Memory injection strategy}
This section discusses the outer stage of our approach, cf. Fig.~\ref{fig:overview}, in which the learner decides \emph{where} to search. 
In particular, a subset of FSCs is selected and passed onto the teacher, as outlined in Sec.~\ref{sec:innerloop}.
We assume access to an abstraction oracle that yields bounds on the value for every state based on an abstraction scheme outlined in Sec.~\ref{sec:abst}.
The learner processes this information and derives a new design space. 
It does so by combining a couple of ingredients:
\begin{inparaenum}
    \item \emph{Adding memory}: By allowing FSCs to store more information, we (drastically) increase the design space. We allow to locally increase memory to keep the growth manageable.
    \item \emph{Removing symmetries}: Similar to \citep{grzes2013isomorph}, it is unnecessary to include symmetric FSCs in the design space.
    \item \emph{Analysing abstractions}: Use the results from the abstraction to guide the search.
\end{inparaenum}


\subsection{Adding Memory}
The key idea of the memory injection strategy is to use the diagnostic information obtained from the preceding inner loop exploring the design space represented by a family $\ffsc$ to construct a new family $\ffsc'$, say. 
These families are based on two fixed memory structures (either as a full or reduced FSC). 
On constructing $\ffsc'$, memory can be added that corresponds to one of the observations, see Sec.~\ref{sec:red_fsc}. This section outlines where to add the memory.

To decide how to extend the family of FSCs, we use the following information: 
\begin{inparaenum}
    \item the maximising policy  $\pi^*$ in MDP $\abst{\ffsc}$ together with its corresponding bounds $ub$\footnote{
Note that the state space of MDP $\abst{\ffsc}$ includes copies of states where memory has been added previously.  Without symmetry reduction, an optimal policy (mostly) takes the same action in these copies, thus making the copies redundant. However, in combination with the symmetry reduction, the outgoing transitions of these copies differ and the copies are then no longer redundant.}, and 
    \item The FSC $\fsc^*$ in $\ffsc$ obtained from the teacher. 
    Such an FSC is not available if the FSC is inadmissible or if the teacher aborts the search. 
\end{inparaenum}
 
If $\fsc^*$ is not available, the memory injection strategy employs  a similar idea as the refinement strategy described in Sec.~\ref{sec:counterexpruning}. 
In particular, it evaluates the significance of the inconsistent observations w.r.t.\ $\pi^{*}$
by aggregating the significance of their inconsistent parameters as defined before.
By adding the memory to the most significant inconsistent observation, we try to resolve the inconsistency and drive the search towards an FSC that is topologically close to $\pi^*$.

If $\fsc^*$ is available, the injection strategy tries to improve it by adding memory to the observations where 
$\fsc^*$ selects a different action than $\pi^*$. Similarly as before, we evaluate the significance of the observations by evaluating the significance of the corresponding parameters. (This is possible as the expected number of visits of states in the MC induced by $F^*$ or $\pi^*$ can be used).
We estimate how changing an action in $\fsc^*$ can improve its value by looking at the state values in $ub$. We use the expected visits in $\fsc^*$ to estimate the impact of these changes. Here, we also consider consistent (but non-trivial) observations in $\pi^*$. 

\vspace{-0.5em}
\subsection{Symmetry reduction} 
Adding memory to a selected observation typically creates FSCs in the resulting family $\ffsc$ that have the same value (see Example~5), often because of some symmetries between the FSCs. Removing these symmetries from $\ffsc$ has two crucial benefits: i) it reduces the size of $\ffsc$ and ii) it reduces the number of inconsistencies of the $\pi^*$ obtained from $\abst{\ffsc}$, see below. The importance of symmetry breaking has been recognised by~\cite{grzes2013isomorph} who proposed a generation strategy for isomorphism-free Moore automata. We propose a different approach based on restricting the family of candidate FSCs.  We first describe where the symmetries are introduced, and then how we avoid this.

Let $\pi^*$ be the optimal inconsistent controller obtained from analysing $\abst{\ffsc}$ where $\ffsc$ describes 1-FSCs (no memory was added) based on $\pi^*$. In line with the above algorithm, we add memory to the inconsistent observation~$z$, where $(n_0,z) \in K$ is the inconsistent parameter with 
$V_{(n_0,z)} = \{(a_1,0), \ldots,  (a_n,0)\}$. This means $\exists {s_1,s_2 \in S},
(a_i,0),(a_j,0)\in V_{(n_0,z)} : \obsFun(s_1)=\obsFun(s_2)=z$ and 
$\pi^*((s_1,n_0)) = (a_i,n_0) \neq (a_j,n_0) = \pi^*((s_2,n_0))$. 
Adding memory creates a new family $\ffsc_n$ where $K$ includes 
the new parameter $(n_1,z)$ having the same domain as $(n_0,z)$. It introduces several symmetric FSCs in $\ffsc_n$. Consider, e.g., $F_0\in \ffsc_n$ where  
 $\gamma(n_0,z)=a_i$, $\gamma(n_1,z)=a_j$ and has a memory mapping $\delta$.
 Then there is also $F_1\in \ffsc_n$ that is symmetric in $z$, i.e. $\gamma(n_0,z)=a_j$, $\gamma(n_1,z)=a_i$, and has a symmetric $\delta$ (otherwise it is the same as $F_0$). 
 Clearly, $F_0$ and $F_1$ are equivalent (i.e., achieve the same value) and thus we want to keep only one of them in $\ffsc_n$. 
 Additionally, $\pi^*$ obtained for $\abst{\ffsc_n}$ has the same inconsistency as before, but now in both parameters.   

The key idea of the symmetry reduction is to restrict $V_{(n_0,z)}$ and $V_{(n_1,z)}$ in $\ffsc_n$ such that symmetric FSCs  are removed and the inconsistency in $\pi^*$ for $\abst{\ffsc_n}$ is decreased while the value of $\pi^*$ is preserved. In this case, the reduction can be easily achieved by removing $a_i$ from $V_{(n_0,z)}$ and $a_j$ from $V_{(n_1,z)}$.
If in some state $s$, $O(s) = z$, the optimal action is $a_i$, then $\pi^*$ will ensure that the predecessors of $s$ use, if possible, the appropriate memory update that leads to $(s,1)$, where this action is available. This decreases the probability of reaching $(s,0)$ as well as the significance of the possible inconsistency associated with $(n_0,z)$.



Removing symmetry for the first memory injection is safe (the optimal solution is preserved). When several memory injections and symmetry reductions are performed, the optimal solution can potentially be removed as the symmetry reduction in one observation may affect the completeness of the reduction in another observation via memory updates. We demonstrate the impact of symmetry reduction in our experimental evaluation.       

\begin{example}
Consider again the maze problem from Example~1 and the specification to minimise the expected number of steps to reach $T$. 
Note that this includes an implicit constraint $\prob_{\geq 1}$ to reach $T$.
Let $\ffsc_1$ be the family of all 1-FSCs.
The inner loops detects that there is no admissible 1-FSC satisfying the constraint: in observation $z_1$ we need to be able to pick both actions $r$ and $l$, and similarly for observation $z_4$. Analysing $\pi^*$ in $\abst{\ffsc_1}$ reveals that the most significant inconsistency is in observation $z_1$.
Thus, memory is added to $z_1$ and the symmetry is removed as follows: $V_{(n_0,z_1)} = \{u,d,r\} \times \{0,1\}$ and $V_{(n_1,z_1)} = \{u,d,l\} \times \{0,1\}$, obtaining a new family~$\ffsc'_1$.
The inner loop again detects that no admissible solution exists, proposing to add memory to observation $z_4$ as well as to break symmetry in actions $u$ and $d$: $V_{(n_0,z_4)} = \{u,r,l\} \times \{0,1\}$ and $V_{(n_1,z_1)} = \{d,r,l\} \times \{0,1\}$.
The third iteration finally yields an optimal FSC with value $7.1\overline{6}$. We emphasise that no additional memory injection can improve upon this.



\end{example}

\vspace{-1em}
\section{Experimental evaluation}
\vspace{-0.5em}
Our evaluation focuses on the following questions:

\noindent
\emph{Q1: How does our approach compare to state-of-the-art approaches to synthesise deterministic FSCs?} 
To this end, we consider the state-of-the-art dual MILP formulation from~\citep{kumar2015history} which uses a max-entropy strategy for adding memory nodes. 
We consider a recent alternative formulation of a primal MILP in  ~\citep{winterer2020strengthening} for treating multi-objective specifications.

\noindent
\emph{Q2: How does our approach compare with state-of-the-art belief-based approaches?} 
Although these approaches share the main idea (i.e., approximate the underlying belief MDP that is prohibitively large or infinite), they approach the problem by constructing an approximation of the belief-MDP. 
We compare with the approach of~\citep{norman2017verification} (implemented in  Prism~\cite{DBLP:conf/cav/KwiatkowskaNP11}) and the recent work of~\citep{bork2022under} (implemented in Storm~\citep{STORM}).
To the best of our knowledge, these methods provide state-of-the-art techniques for finding policies in belief MDPs for indefinite-horizon specifications, i.e., without discounting.  

\noindent
\emph{Q3: What is the effect of our heuristics on the run-time and the value of the resulting FSCs?}  

\paragraph{Selected benchmarks and experimental setting} 
A direct comparison with MILP-based approaches is complicated as they mostly consider other specifications (discounted rewards) and  stochastic observations.
Therefore, we selected the \emph{Hallway} model from~\citep{kumar2015history} 
and manually translated it to an almost equivalent model~\footnote{The values of the resulting FSCs are comparable.}. 
We also took a multi-objective variant of a \emph{4×4grid avoid} model  from~\citep{winterer2020strengthening} (denoted as \emph{Grid-av 4.0}) that enables a direct comparison with the multi-objective MILP optimisation. 
The main evaluation considers representative benchmarks from \citep{bork2020verification,bork2022under} extended by a few more involved variants~\footnote{The results with STORM and PRISM were provided by the authors. The experiments run on a comparable machine.}. 
Table~\ref{tab:stats} lists the statistics of the models. Our results run on a single core of a machine equipped with an Intel i5-8300H CPU and 24 GB of RAM.

 \begin{table*}[t]
  \centering
 \renewcommand{\arraystretch}{0.8}
 \setlength{\tabcolsep}{3pt}
 \scalebox{0.95}{
\begin{tabular}{|c|rrr|r||c|rrr|r||c|rrr|r|}
\hline
Model & $S$ & $\Act$ & $\Obs$ & Spec. & Model & $S$ & $\Act$ & $\Obs$ & Spec. & Model & $S$ & $\Act$ & $\Obs$ & Spec. \\ \hline 
 
 Grid-av 4-0 & 17 & 59 & 4 & $P_{\max}$ &  Grid-av 4-0.1 & 17 & 59 & 4 & $P_{\max}$ &  Grid 30-sl & 900 & 3587 & 37 & $P_{\max}$\\ 

 Maze sl & 15 & 54 & 3 & $R_{\min}$ & Crypt 4 & 1972 & 4612 & 510 & $P_{\max}$ &   Nrp 8 & 125 & 161 & 41 & $P_{\max}$
 \\ 
  
    Hallway & 61 & 301 & 23 & $P_{\max}$ & Drone 4-1 & 1226 & 3026 & 384 & $P_{\max}$ & Drone 4-2 & 1226 & 3026 & 761 & $P_{\max}$ \\

    Refuel 6 & 208 & 565 & 50 & $P_{\max}$ & Netw-p  2-8-20 & $3\!\cdot\!10^4$ & $3\!\cdot\!10^4$ & 4909 &  $R_{\max}$ &    Rocks 12 & 6553 & $3\!\cdot\!10^4$ & 1645 & $R_\mathrm{min}$ \\ \hline

\end{tabular}
}
 \caption{The statistics of the twelve POMDP models and the optimisation objective.}
 \vspace{-0.5em}
 \label{tab:stats}
   \end{table*}

\vspace{-0.5em}

\subsubsection*{Q1: Comparison to MILP-based FSC synthesis} 
We first compare our approach with~\citep{kumar2015history} on the \emph{Hallway} model. The dual MILP optimisation for the fixed-size reactive FSC (equivalent to our 1-FSC) achieved the value 0.32 in 900s.
This corresponds to an FSC where the expected number of steps to reach the target equals 22.2\footnote{\citep{kumar2015history} use a discount factor of 0.95.} 
Using the memory injection strategy, they found an FSC with 14 additional memory nodes in 3345s achieving value 0.46, i.e., 15.1 expected steps. 
Our complete strategy explored all 1-FSCs in less than 1s and found a solution achieving 18.5 expected steps. 
The restricted exploration of full 3-FSCs found a solution achieving 14.9 expected steps in 156s. 
The default strategy used in Tab~\ref{tab:main} (see below) added one memory node and found a solution achieving 16.1 expected steps in less than 1s.
Although the value of the resulting FSCs cannot be exactly compared, \emph{these results clearly demonstrate that our approach is superior to MILP-based synthesis methods.} 

We also compare our approach with~\citep{winterer2020strengthening} on the \emph{Grid-av 4} model  with a constraint on the reachability probability and the minimisation of an reward. The best solution of the MILP optimisation with a restricted randomisation and memory injection has value 3.43 (found in 2.8s). 
This solution is obtained by our default strategy within 1s by adding one memory node. 
In 389s, it added nine memory nodes and found a solution with value 3.29. 
This shows that \emph{our inductive approach outperforms the MILP optimisation also on multi-objective specifications}.

 \begin{table}[t]
 \centering
 \renewcommand{\arraystretch}{0.8}
 \setlength{\tabcolsep}{2.5pt}
 \scalebox{0.95}{
 \begin{tabular}{|c||r||r|r||r|r|}
\hline
\multicolumn{1}{|c||}{Benchmark}  & \multicolumn{1}{c||}{PRISM} &  \multicolumn{2}{c||}{STORM}&   \multicolumn{2}{c|}{Inductive synthesis} \\

\multicolumn{1}{|c||}{model}  & 
\multicolumn{1}{c||}{}& \multicolumn{1}{c|}{first} & \multicolumn{1}{c||}{best} & \multicolumn{1}{c|}{fastest} & \multicolumn{1}{c|}{best} \\ \hline \hline

Grid-av  & 0.21 & $ 0.86$ & $ 0.93$ & \textbf{0.93} (3) & 0.93(4)$^\dagger$  \\ 
4-0 &     5s & < 1 s & 14s   & \textbf{<1s} & <1s  \\ \hline

Grid-av  & 0.21 & $0.82$ & $0.85$ & 0.85 (2) & \textbf{0.92 (10)} \\
4-0.1 &    1s & < 1 s & 1913s   & < 1s  & \textbf{874s}  \\ \hline



Grid    &   \multirow{2}{*}{MO/TO}  & 121 & - & TO & \textbf{119(6)}  \\ 
30-sl   &   & <1s & -  & TO &  $\mathbf{3s^*}$  \\ \hline

Maze    &  \multirow{2}{*}{SE}  &  7.68 & - & \textbf{7.14}(3) & \textbf{7.09(3f)}  \\ 
sl         &   & 1s & -  & $\mathbf{<1 s}$ & $\mathbf{18s^*}$   \\ \hline

Crypt &  0.33 & \textbf{0.33} &  - & 0.33(0) &- \\ 
4    &   20s & \textbf{< 1s} & -  & 3s & - \\ \hline 

Nrp  &  0.13 &  0.12 & - & \textbf{0.13(0)}& -  \\ 
8   &    2s &< 1s & -  & \textbf{< 1s} & -   \\ \hline

Hallway  &  \multirow{2}{*}{SE} &  19.2 & 19.3 & \textbf{16.3(1)}&  \textbf{14.9(3f)}  \\ 
         &   & < 1s & 3s  & \textbf{< 1s} & $\mathbf{156s^*}$   \\ \hline
         
Drone &    \multirow{2}{*}{MO/TO} & $ 0.79$ & - & 0.79 (0) & \textbf{0.87(0)}   \\ 
4-1 &      & < 1 s & -  &  12 s & \textbf{120s}   \\ \hline

Drone &   \multirow{2}{*}{MO/TO} & $ 0.86$ & $ 0.92$ & 0.93 (0) & \textbf{0.97(2)}   \\ 
4-2 &     & < 1 s & 1902s  &  3 s & \textbf{611s}  \\ \hline

Refuel&  0.67 & $\mathbf{ 0.67}$ & $ 0.67$ & 0.63(8) & 0.67(2f)\\ 
6 &     4625s & \textbf{< 1 s} & 2076s  & 2.6s  & $170s^*$  \\ \hline

Netw-p   & $\mathbf{557}$ & $537$ & - & 539(0) & -  \\ 
2-8-20     &  \textbf{2355s} & 2s & -   & 210s  & -  \\ \hline


Rocks  &  \multirow{2}{*}{MO/TO} & $0.38$ & $\mathbf{ 20}$ & 42(0) & -  \\ 
12 &     & 1 s & \textbf{230s}   & 3s  & -  \\ \hline

\end{tabular}
}
\caption{The comparison with belief-base methods. Bold entries denote the best solutions, -- indicates that no better solution was found within 30 minutes, * indicates that the result was found using a non-default strategy, MO/TO  denotes out of memory or timeout,
and SE indicates errors.}
\label{tab:main}
\vspace{-0.5em}
  \end{table}

\subsubsection*{Q2: Comparison to belief-based methods}


Table~\ref{tab:main} summarises key experimental results related to \textbf{Q2}. 
The columns list the following information (from left to right): the model and its version, the lower bounds provided by~\citep{norman2017verification}  and its run-time, the lower bounds provided by~\citep{bork2022under} and its run-time (for two settings: the fastest synthesis and the best bound), the results provided by our approach (including the number of added memory nodes) and its run-time (the first interesting solution and the best solution found).

To simplify the presentation, this table shows results (except for the entries denoted by $^*$) achieved by our approach using the default setting: the inner loop is instantiated by the pure MDP abstraction oracle with the incomplete refinement strategy, and the outer loop uses the memory injection strategy with symmetry reduction. 
The impact of our optimisation heuristics as well as the results for multi-objective specifications are discussed under \textbf{Q3}.

The results clearly demonstrate that \emph{our inductive approach is highly competitive with the belief-state space approximation for indefinite-horizon specifications}. 
For the models having moderate number of observation/actions, we provide better trade-offs between the run-time and the values of the found policies.
Moreover, we found small FSCs that improve the lower-bounds in~\citep{bork2022under}.  
For models with a large number of observations/actions, we found small high-quality FSCs in comparable run-time. 
For the \emph{Rocks} model and a larger \emph{Netw} model, we failed to find a good solution.
We highlight two interesting results: 
For \emph{Grid-av 4-0}, our strategy injected four memory nodes (see $\dagger$) and achieved the bound provided by the symmetry-free MDP abstraction which guarantees the global optimum.
For \emph{Drone 4-2}, we found a very small FSC that implements the known upper bound on the solution value~\citep{bork2020verification}. 
  
\vspace{-0.3em}
\subsubsection*{Q3: The effect of optimisation heuristics}
\emph{Efficacious heuristics}
We generally remark that the design spaces in this paper are several orders of magnitude bigger than the design spaces supported by the more general-purpose inductive synthesis framework in ~\citep{DBLP:conf/cav/AndriushchenkoC21}. The differences can mostly be explained by the tailored representation and novel heuristics. 

\emph{Symmetry reduction:} This is quite beneficial as it reduces the design space and, more importantly, considerably helps the memory injection strategy correctly select the most promising observation. 
For example, in the \emph{Maze2} model, the memory injection without symmetry reduction repeatedly adds memory only due to a single observation and the optimal solution is not found. 
On the other hand, symmetry reduction can discard an optimal solution as demonstrated in the \emph{Refuel 6, Hallway} and \emph{Grid large} models. For these models, the last column of Tab.~\ref{tab:main} lists the results of the incomplete exploration of full 2-FSCs and 3-FSC denoted as 2f and 3f, respectively.

\emph{Incomplete search:} the incomplete refinement strategy and CE generalisation is essential for handling large number of observations/actions. The complete exploration, e.g., fails to find a good solution for the \emph{Drone} models. In our experiments, we did not observe that the incomplete exploration discards import solutions except the \emph{Grid-av \mbox{4-0.1}} model, where the full exploration found in 580s (after 9 memory injections) a better solution achieving 0.93. 

\emph{Hybrid approach:} For the models in Table~\ref{tab:main}, the hybrid approach (combining the exploration using the MDP abstraction and CE pruning) does not improve the synthesis process. 
However, for models where the MDP abstraction is significantly larger than the induced MCs corresponding to the candidate FSCs, the hybrid approach is superior, as exemplified by the \emph{Grid 30-sl} model, a more complex variant of the grid-like model, where the MDP abstraction is 15x larger.
For this model, the standard settings do not find an admissible FSC within 30 minutes. Using the hybrid approach in the inner loop, we found an FSC improving the solution found by the belief-based method within 3s.

\emph{Multi-objective (MO) specifications:} Apart from the MO variant of the \emph{Grid-av 4-0} model discussed in \textbf{Q1}, we also considered a MO variant of the \emph{Maze sl} model including a  more complicated specification with an additional reach-avoid constraint. The constraint restricts the optimal FSC, but the run-time of the synthesis remains < 1s.  


\vspace{-0.5em}
\section{Conclusion}
\vspace{-0.5em}
This paper presents a first inductive-synthesis based framework for finding finite-state controllers (FSCs) in POMDPs. Key ingredients are the novel heuristics to incrementally construct the memory structure of the FSC as well as two oracles for searching and evaluating families of FSCs. The experimental results show promising results indicating that this framework is competitive with the state-of-the-art alternatives. Future work includes the integration of belief-based approaches as an additional oracle.

\clearpage
\pagebreak
\bibliography{main}

\end{document}